\documentclass[showpacs,preprintnumbers,amsmath,amssymb]{revtex4}
\usepackage{mathrsfs}
\usepackage{graphicx}

\begin{document}

\title{Statefinder Parameters for Tachyon Dark Energy Model}

\author{Ying Shao\footnote{sybb37@student.dlut.edu.cn},
Yuanxing Gui\footnote{Corresponding author: guiyx@dlut.edu.cn}}

\address{School of Physics \& Optoelectronic Technology, Dalian
University of Technology, Dalian, 116024, P. R. China}


\begin{abstract}

In this paper we study the statefinder parameters for the tachyon
dark energy model. There are two kinds of stable attractor solutions
in this model. The statefinder diagrams characterize the properties
of the tachyon dark energy model. Our results show that the evolving
trajectories of the attractor solutions lie in the total region and
pass through the LCDM fixed point, which is different from other
dark energy model.
\end{abstract}

\maketitle
\section{Introduction}

Current astrophysical observations indicate that our universe has
entered a phase of accelerated expansion in the recent past.$^{1,2}$
The accelerated expansion has been attributed to the existence of
mysterious dark energy with negative pressure. The Wilkinson
Microwave Anisotropy Probe (WMAP) satellite experiment tells us that
the usual baryonic matter, dark matter and dark energy occupy about
4\%, 23\% and 73\% of the total energy of the universe,
respectively. At present, candidates for dark energy have been
widely studied. With the exclusion of the cosmological constant
$\Lambda$, most dark energy modelling using scalar field has
followed the quintessence$^{3,4}$ paradigm of a slowly rolling
canonical scalar field. However, there has been increasing interest
in loosening the assumption of a canonical kinetic term. In its most
general form, this idea is known as k-essence.$^{5}$ A more specific
choice is the tachyon,$^{6}$ which can be viewed as a special case
of k-essence models with Dirac-Born-Infeld (DBI) action.$^{7}$ This
kind of scalar field is motivated by string theory as the
negative-mass mode of the open string perturbative spectrum.
Sen$^{6}$ showed that the decay of D-branes produces a pressureless
gas with finite energy density that resembles classical dust. A
rolling tachyon has an interesting equation of state whose parameter
smoothly interpolates between $-1$ and $0$. This has led to flurry
of attempts being made to construct viable candidate for the
inflaton at high energy. Meanwhile the tachyon can also act as a
source of dark energy depending on the form of tachyon potential.
Tachyon dark energy has been explored by many authors, for example
Refs [8,9,10]. One of these studies is to investigate whether there
are any distinctive signatures of non-canonical actions available to
be probed by observations.

Since there are more and more models proposed to explain the cosmic
acceleration, it is very desirable to find a way to discriminate
between the various contenders in a model independent manner. In Ref
[11], Sahni et al proposed a cosmological diagnostic pair $\{r,s\}$
called statefinder, which is defined as
\begin{equation}
r\equiv\frac{\dddot{a}}{aH^3},~~~~s\equiv\frac{r-1}{3(q-1/2)}.\label{eq:statefinder}
\end{equation}
to differentiate among different forms of dark energy. Here $q$ is
the deceleration parameter. The statefinder is a geometrical
diagnostic that it depends on the scalar factor $a$. Since different
cosmological models involving dark energy exhibit qualitatively
different evolution trajectories in the $s-r$ plane, this
statefinder diagnostic can differentiate these dark energy models.

In this paper, we will study the statefinder parameters for the
tachyon field dark energy model. It is found that the evolving
trajectories of this model are different from other dark energy
models. The paper is organized as follows: In Sec. 2, we specify the
form of tachyonic potential and obtain the autonomous equations of
tachyon dark energy model. In Sec. 3, it is analyzed to the evolving
trajectories of this model in the statefinder parameter plane.
Section 4 is our conclusions.

\section{the tachyon dark energy model}

The evolution equations for a flat FRW cosmological model filled
with a tachyon field T and a barotropic perfect fluid with equation
of state $p_\gamma=(\gamma-1)\rho_\gamma$ are
\begin{eqnarray}
\frac{\ddot{T}}{1-\dot{T}^2}&+&3H\dot{T}+\frac{1}{V(T)}\frac{dV(T)}{dT}=0,\\
\dot{\rho}_\gamma&+&3\gamma H\rho_\gamma=0,
\end{eqnarray}
which are subject to the Friedmann equation
\begin{equation}
3H^2=\frac{V(T)}{\sqrt{1-\dot{T}^2}}+\rho_\gamma,
\end{equation}
where dots denote differentiation with respect to cosmic time $t$
and $H$ is the Hubble parameter.

The density $\rho_T$ and pressure $p_T$ of tachyon field are given
by, then
\begin{eqnarray}
\rho_T&=&\frac{V(T)}{\sqrt{1-\dot{T}^2}},\\
p_T&=&-V(T)\sqrt{1-\dot{T}^2}.
\end{eqnarray}

We introduce the following dimensionless quantities:
\begin{equation}
x\equiv\dot{T},~~~~y\equiv\frac{V(T)}{3H^2},~~~~z\equiv\frac{\rho_\gamma}{3H^2}.
\end{equation}
Thus the fractional densities of the two fluids are defined as,
respectively
\begin{eqnarray}
\Omega_T &\equiv& \frac{\rho_T}{3H^2}=\frac{y}{\sqrt{1-x^2}},\\
\Omega_\gamma &\equiv& \frac{\rho_\gamma}{3H^2}=z.
\end{eqnarray}

Consider the inverse square potential$^{12}$ $V(T)=\beta T^{-2}$
now. We can cast the evolution equations in the following autonomous
form:
\begin{eqnarray}
\frac{dx}{dN}&=&3(x^2-1)(x-\sqrt{\alpha y}),\\
\frac{dy}{dN}&=&3y[x(x-\sqrt{\alpha y})+z(\gamma-x^2)],\\
\frac{dz}{dN}&=&3z(z-1)(\gamma-x^2),
\end{eqnarray}
where the number of e-folds $N\equiv\ln a$ and the constant
$\alpha\equiv 4/3\beta>0$. From the definitions of these new
variables, we obtain the equation $\omega_T$ of state of tachyon
field
\begin{equation}
\omega_T=\frac{p_T}{\rho_T}=x^2-1.
\end{equation}
\section{the statefinder parameters for tachyon dark energy model}

In order to discriminate between different forms of dark energy,
Sahni et al$^{11}$ proposed a geometrical diagnostic
method---statefinder papameter. Since different cosmological models
involving dark energy exhibit qualitatively different evolution
trajectories in the $s-r$ plane, this statefinder diagnostic can
differentiate various kinds of dark energy models. By far some
models, including the cosmological constant, quintessence, phantom,
quintom, the Chaplygin gas, holographic models, interacting and
coupling dark energy models$^{11,14-20}$ have been successfully
differentiated. For example, the statefinder parameters correspond
to a fixed point $\{r=1,s=0\}$ for the spatially flat LCDM
cosmological model which is a good fit for the
observation$^{21,22}$, the quintessence model with inverse power law
potential, the phantom model with power law potential and the
chaplygin gas model all tend to approach the LCDM fixed point, but
for quintessence and phantom models the trajectories lie in the
regions $s>0,r<1$ while for Chaplygin gas model the trajectories lie
in the regions $s<0,r>1$. In this paper, we apply the statefinder
diagnostic to the tachyon field model. To begin with, statefinder
parameters (\ref{eq:statefinder}) can be rewritten as
\begin{eqnarray}
r&=&1+\frac{9}{2}\gamma^2-\frac{9}{2}\gamma+\frac{1}{2}(9\gamma-9\gamma^2)\Omega_T+\frac{9}{2}\omega_T(1+\omega_T)\Omega_T-\frac{3}{2}\omega'_T\Omega_T,\\
s&=&\frac{(3\gamma^2-3\gamma)(1-\Omega_T)+3\Omega_T\omega_T(1+\omega_T)-\Omega_T\omega'_T}{3(\gamma-1)(1-\Omega_T)+3\Omega_T\omega_T},
\end{eqnarray}
which $\omega'_T=d\omega_T/dN$, and the deceleration parameter is
also given
\begin{equation}
q=\frac{3\gamma}{2}-1-\frac{3}{2}(\gamma-1)\Omega_T+\frac{3}{2}\omega_T\Omega_T.
\end{equation}

In the following we will discuss the statefinder for the attractor
solutions in Ref [13]: tachyon dominated solution and tracking
solution. Firstly, we discuss the attractor solution $P$ at
$(x,y)=(\sqrt{\alpha y_1},y_1)$
($y_1\equiv\frac{\sqrt{\alpha^2+4}-\alpha}{2}$), which is an
asymptotically stable mode and corresponds to tachyon dominated
solutions for $\gamma>\gamma_1$ ($\gamma_1\equiv\alpha y_1$).
Nextly, there exists another attractor solution Q at
$(x,y)=(\sqrt{\gamma},\frac{\gamma}{\alpha})$. When
$0\leq\gamma<\gamma_1$, $Q$ is always asymptotically stable and
corresponds to the tracking solution. In Ref [13] the constant
$\alpha=1.5$, we can obtain $y_1=0.5$ and $\gamma_1=0.75$. With the
condition of the above attractor solutions, we consider the
constants $\gamma=4/3$ for the tachyon dominant solution and
$\gamma=0.3$ for the tracking solution. In addition, the two special
values $\gamma=0$ and 1 are also studied. In Fig. 1 and Fig. 2 we
show the time evolution of the statefinder parameter pairs $\{r,s\}$
and $\{r,q\}$. The plot is for variable interval $N\in[-5,10]$. We
see clearly that the curves pass through the LCDM fixed point
$\{0,1\}$. And the evolving trajectories lie in the total region,
which is different from other quintessence and phantom models. In
Fig. 3 and Fig. 4 we plot the evolution trajectories of the
statefinder parameters versus $N$ diagram of the stable attractor
solutions. It is noted that $\gamma$ causes the deviation between
statefinder parameters and LCDM scenario. In the figure of $N-r$,
the curve for $\gamma=0$ approaches to LCDM scenario after $N=0$,
while for other cases the distance from the curves to LCDM scenario
becomes large with the decreasing of $\gamma$. In the figure of
$N-s$, the constant $\gamma$ also causes the deviation between the
curves and LCDM scenario. In Fig. 5 the equation of state for the
tachyon field has been shown. The equations of state for $\gamma=0$
and $\gamma=0.3$ tend to $-1$ and $-0.7$ after oscillations,
respectively, while for the other two cases $\omega_T$ would violate
the basic condition $\omega<-1/3$, which can not lead to the
acceleration. So, through the statefinder diagnostic, we not only
characterize the properties of the tachyon dark energy model, but
also show the difference from the other dark energy models.
\begin{figure}
\centering\includegraphics[width=3.0in,height=1.85in]{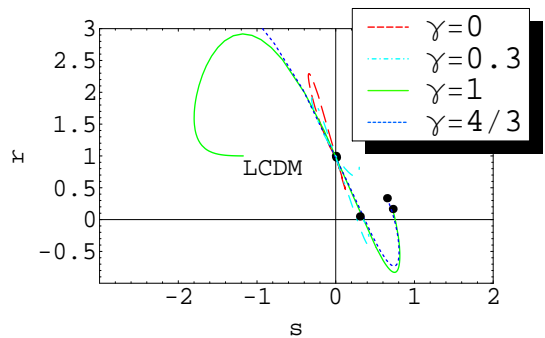}
\caption{The figure is $s-r$ diagram of the stable attractor
solutions. The curves evolve in the variable interval $N\in[-5,10]$.
Selected curves for $\alpha=1.5$, $\gamma=0$, $\gamma=0.3$,
$\gamma=1$ and $\gamma=4/3$ respectively. Dots locate the current
values of the statefinder parameters. Here, since the current values
of the statefinder parameters for $\gamma=0$  are almost equal to
the LCDM fixed point $\{0,1\}$, dots for the two cases
coincide.}\label{rs}
\end{figure}

\begin{figure}
\centering\includegraphics[width=3.0in,height=1.85in]{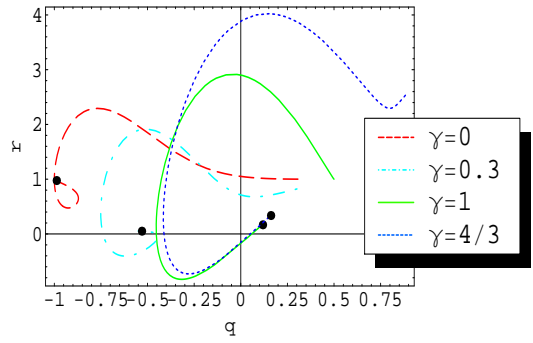}
\caption{The figure is $q-r$ diagram of the stable attractor
solutions. The curves evolve in the variable interval $N\in[-5,10]$.
Selected curves for $\alpha=1.5$, $\gamma=0$, $\gamma=0.3$,
$\gamma=1$ and $\gamma=4/3$ respectively. Dots locate the current
values of the statefinder parameters.}\label{rq}
\end{figure}

\begin{figure}
\centering\includegraphics[width=3.0in,height=1.85in]{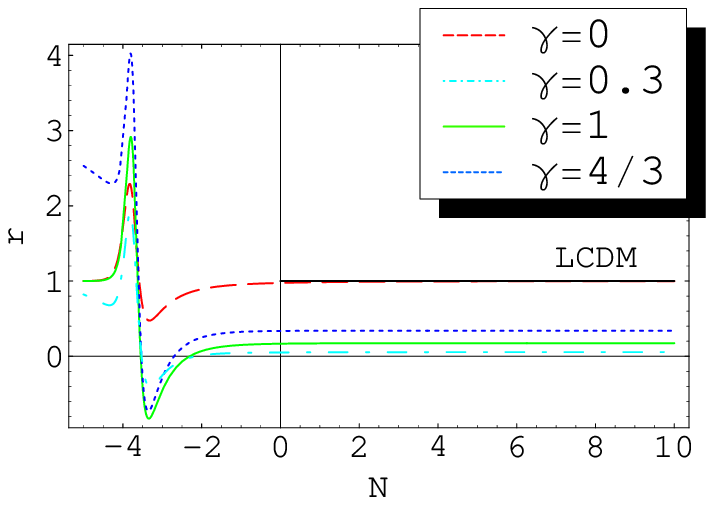}
\caption{The figure is $N-r$ diagram of the stable attractor
solutions in the variable interval $N\in[-5,10]$. Selected curves
for $\alpha=1.5$, $\gamma=0$, $\gamma=0.3$, $\gamma=1$ and
$\gamma=4/3$ respectively.}\label{rN}
\end{figure}

\begin{figure}
\centering\includegraphics[width=3.0in,height=1.85in]{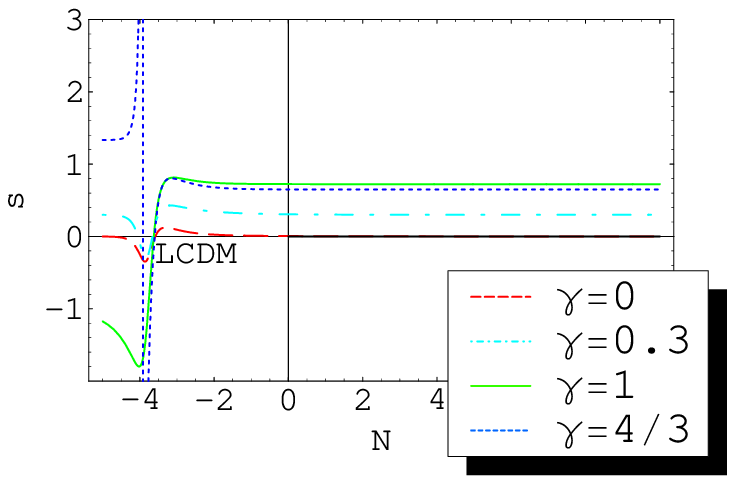}
\caption{The figure is $N-s$ diagram of the stable attractor
solutions in the variable interval $N\in[-5,10]$. Selected curves
for $\alpha=1.5$, $\gamma=0$, $\gamma=0.3$, $\gamma=1$ and
$\gamma=4/3$ respectively.}\label{sN}
\end{figure}

\begin{figure}
\centering\includegraphics[width=3.0in,height=1.85in]{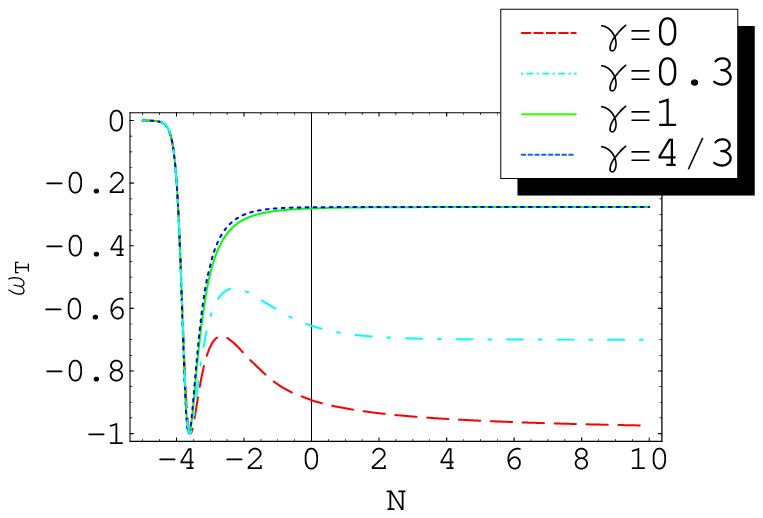}
\caption{The figure is the equation of state for the tachyon field
$\omega_T$ versus $N$. The curves evolve in the variable
$N\in[-5,10]$. Selected curves for $\alpha=1.5$, $\gamma=0$,
$\gamma=0.3$, $\gamma=1$ and $\gamma=4/3$ respectively.}\label{wN}
\end{figure}

\section{conclusions}

In this paper, we study the statefinder parameters of the tachyon
dark energy model. We analyze two cases of the stable attractor
solutions: tachyon dominated solution and tracking solution. The
statefinder diagrams characterize the properties of the tachyon dark
energy model. Our results show that the evolving trajectories of
this model lie in the total region and pass through the LCDM fixed
point. The statefinder diagnostic can differentiate the tachyon
model from other dark energy models. We hope that the future high
precision observation will be capable of determining these
statefinder parameters and consequently shed light on the nature of
dark energy.

\section*{Acknowledgments}

This work was supported by National Science Foundation of China
under Grant NO.10573004.

\end{document}